\documentclass[referee]{cjaa}           % referee version: for submission

\usepackage{graphicx}                   %for PS/EPS graphics inclusion, new
\input{epsf.sty}                        %for PS/EPS graphics inclusion, old
\input{psfig.sty}                       %for PS/EPS graphics inclusion, old

\begin{document}

   \title{BBN, ADTree and MLP Comparison in Separating Quasars from Large Survey Catalogues\,$^*$
%\footnotetext{$*$ Supported by the National Natural Science Foundation of China.}
}

   \volnopage{Vol.0 (200x) No.0, 000--000}      %%preserved for Editor. DOn't remove!
   \setcounter{page}{1}          %%starting page, preserved for Editor. DOn't remove!

   \author{Yanxia Zhang
      \inst{1}\mailto{}
   \and Yongheng Zhao
      \inst{1}
      }
   \offprints{Y. Zhang}
   \institute{National Astronomical Observatories, Chinese Academy of Sciences,
             Beijing 100012, China\\
             \email{zyx@bao.ac.cn}}

   \date{Received~~2006.6; accepted~~}

   \abstract{
We compare the performance of Bayesian Belief Networks (BBN),
Multilayer Perceptron (MLP) networks and Alternating Decision
Trees (ADtree) on separating quasars from stars with the database
from the 2MASS and FIRST survey catalogs. Having a training sample
of sources of known object types, the classifiers are trained to
separate quasars from stars. By the statistical properties of the
sample, the features important for classification are selected. We
compare the classification results with and without feature
selection. Experiments show that the results with feature
selection are better than those without feature selection. From
the high accuracy, it is concluded that these automated methods
are robust and effective to classify point sources, moreover they
all may be applied for large survey projects (e.g. selecting input
catalogs) and for other astronomical issues, such as the parameter
measurement of stars and the redshift estimation of galaxies and
quasars.  }

   \authorrunning{Y. Zhang \& Y. Zhao }            %author_head in even pages
   \titlerunning{BBN, ADTree and MLP Comparison}  % title_head in odd pages

   \maketitle
\section{Introduction}

The rapid emergence of huge, uniform, multivariate databases from
specialized survey projects and telescopes has lead to the coming
of the `information age' in astronomy, just like the `data
avalanche' faced in other fields. Powerful database systems for
collecting and managing data are in use in virtually all large and
mid-range astronomical institutes. How to collect, save, organize,
and mine the data efficiently and effectively is an important
problem. Due to the large size of the databases, it is impossible
to manually analyze the data for knowledge discovery. Therefore
the automated extraction of useful knowledge from huge amounts of
data is widely recognized now, and leads to a rapidly developing
market of automated analysis and discovery tools. Data mining and
knowledge discovery are techniques to identify valid, novel,
potentially useful, and ultimately understandable patterns hidden
in very large databases. Automated discovery tools can provide the
potential and advantages to mine the raw data and obtain the
extracted high level information to the analyst or astronomers.

Just like statistics, the diversity of tasks and techniques in
data mining is broad. For example, Fayyad et al. (1996) divided
data mining tasks into six flavors: (i) classification; (ii)
regression; (iii) clustering; (iv) summarization; (v) dependency
modelling (structural and quantitative); and (vi) change modelling
(changes from previous or normative values). In astronomy, the
automatic classification of objects from catalogues is a common
issue encountered in many surveys (Zhang \& Zhao 2003; Zhang \&
Zhao 2004;  Zhang et~al. 2004). From a list of values of variables
associated with a celestial object, it is desired to identify the
object's type (eg. star, galaxy). In this paper we apply three
automated methods: Bayesian Belief Networks (BBN), Multilayer
Perceptron (MLP) networks and Alternating Decision Trees (ADtree)
to classify objects as quasars or non-quasars using the
cross-matched results of a radio survey and a near infrared
survey. Such classification is helpful to preselect quasar
candidates for large survey projects.

The structure of this paper is as follows: Section 2 presents the
data collection and attribute selection. Section 3 gives a brief
introduction of BBN, MLP and ADTree. The procedure to get the
classifiers and the classification results are presented in
section 4. Section 5 summarizes and concludes the present work.

\section{Data Sample and Chosen Attributes}
We describe here near infrared, radio and known catalogs as
follows. Table 1 summarizes the characteristics of the two
surveys.

The Two Micron All Sky Survey (2MASS) project (Cutri et~al. 2003)
is designed to close the gap between our current technical
capability and our knowledge of the near-infrared sky. 2MASS uses
two new, highly-automated 1.3-m telescopes, one at Mt. Hopkins,
AZ, and one at CTIO, Chile. Each telescope is equipped with a
three-channel camera, each channel consisting of a 256x256 array
of HgCdTe detectors, capable of observing the sky simultaneously
at $j$ (1.25\,$\mu$m), $h$ (1.65\,$\mu$m), and $k$ (2.17\,$\mu$m),
to a 3$\sigma$ limiting sensitivity of 17.1, 16.4 and 15.3\,mag in
the three bands. The number of 2MASS point sources adds up to
470,992,970.

The Faint Images of the Radio Sky at Twenty centimeters (FIRST)
began in 1993. It uses the VLA (Very Large Array, a facility of
the National Radio Observatory (NRAO)) at a frequency of 1.4\,GHz,
and it is slated to 10,000\,deg$^2$ of the North and South
Galactic Caps, to a sensitivity of about 1\,mJy with an angular
resolution of about 5\,arcsec. The images produced by an automated
mapping pipeline have pixels of 1.8\,arcsec, a typical rms of
0.15\,mJy, and a resolution of 5\,arcsec; the images are available
on the Internet (see the FIRST home page at
http://sundog.stsci.edu/ for details). The source catalogue is
derived from the images. A new catalog (Becker et al. 2003) of the
FIRST Survey has been released that includes all taken from 1993
through September 2002, and contains about 811,000 sources
covering 8422\,deg$^2$ in the North Galactic cap and 611\,deg$^2$
in the South Galactic cap. The new catalog and images are
accessible via the FIRST Search Engine and the FIRST Cutout
Server.

The 12th edition catalogue of quasars and active nuclei (Cat.
VII/248, V\'eron-Cetty \& V\'eron 2006) is an update of the
previous versions, which now contains 85221 quasars, 1122 BL Lac
objects and 21737 active galaxies (including 9628 Seyfert 1s),
almost doubling the number listed in the 11th edition. As in the
previous editions no information about absorption lines of X-ray
properties are given, but absolute magnitudes are given, assuming
H$_{0}=50$\,km/s/Mpc and q$_0=0$. In this edition the 20\,cm radio
flux is listed when available, in place of the 11\,cm flux.

\begin{table*}[ht]
\begin{center}
\caption{SUMMARY OF CATALOG CHARACTERISTICS}
\begin{tabular}{llllll}
\hline \hline
      &            &           &Resolution&&\\
Survey& Wavelength &Sensitivity&(arcsec)&Number of Souces&Coverage Area\\
\hline
 FIRST &21\,cm     &1\,mJy       &5& 811,000 &9033\,deg$^2$\\
 2MASS &$j$(1.25\,$\mu$m)&15.8\,mag$^a$&0.5& 470,992,970 &41252.96\,deg$^2$\\
       &$h$(1.65\,$\mu$m)&15.1\,mag$^a$\\
       &$k$(2.17\,$\mu$m) &14.3\,mag$^a$\\
\hline
 $^a$For S/N$=10$. \\
\hline
\end{tabular}
\bigskip
\end{center}
\end{table*}

The Tycho-2 Catalogue (Cat. I/259, Hog et al. 2000) is an
astrometric reference catalogue containing positions and proper
motions as well as two-color photometric data for the 2.5 million
brightest stars in the sky. The Tycho-2 positions and magnitudes
are based on precisely the same observations as the original Tycho
Catalogue (hereafter Tycho-1; see Cat. I/239) collected by the
star mapper of the ESA Hipparcos satellite, but Tycho-2 is much
bigger and slightly more precise, owing to a more advanced
reduction technique. Components of double stars with separations
down to 0.8\,arcsec are included. Proper motions precise to about
2.5\,mas/yr are given.

We firstly positionally cross-matched the 2MASS catalogue with the
FIRST catalogue within 5 arcsecond radius, then crossed out the
one-to-many entries and got 153135 one-to-one entries. Secondly
the entries were cross-matched with qso.dat of the V\'eron-Cetty
\& V\'eron 2006 catalog and the Tycho-2 catalog within 5 arcsecond
radius, respectively. Similarly not considering the one-to-many
entries, we obtained 2389 quasars and 1353 stars from the 2MASS
and FIRST catalogues. The chosen attributes from different bands
are $logFpeak$ ($Fpeak$: peak flux density at 1.4\,GHz), $logFint$
($Fint$: integrated flux density at 1.4\,GHz), $fmaj$ (fitted
major axis before deconvolution), $fmin$ (fitted minor axis before
deconvolution), $fpa$ (fitted position angle before
deconvolution), $j-h$ (near infrared index), $h-k$ (near infrared
index), $k+2.5logFint$, $k+2.5logFpeak$, $j+2.5logFpeak$,
$j+2.5logFint$. To see the statistical properties of this sample,
the mean values of parameters are listed in Table 2. Meanwhile the
distributions of parameters are shown in Fig.1. As shown by Table
2, some mean values have rather large scatters. The values of
$logFpeak$, $logFint$, $k+2.5logFint$, $k+2.5logFpeak$,
$j+2.5logFpeak$, $j+2.5logFint$ for quasars are obviously bigger
than those of stars. This means that quasars are generally
stronger radio emitters than stars. In addition, the values of
$j-h$ and $h-k$ of quasars are larger than those of stars, i.e.
quasars are redder than stars. Moreover Table 1 and Fig.1 indicate
that $fmaj$, $fmin$ and $fpa$ are unimportant to discriminate
quasars from stars while other attributes are useful. Therefore in
the following we classify quasars from stars considering two
situations: the sample 1 (S1) with all attributes and the sample 2
(S2) without $fmaj$, $fmin$ and $fpa$.

\begin{table*}[ht]
\begin{center}
\caption{The mean values of parameters for the samples}
\begin{tabular}{llllll}
\hline \hline
Parameters   &  stars    &  quasars  \\
\hline
$logFpeak$ &0.46$\pm$ 0.46&1.12$\pm$ 0.87\\
$logFint$  &0.55$\pm$ 0.49&1.18$\pm$ 0.91\\
$fmaj$       &7.22$\pm$ 2.95&6.76$\pm$ 2.93\\
$fmin$       &5.51$\pm$ 1.28&5.51$\pm$ 1.16\\
$fpa$        &92.16$\pm$ 59.29&87.61$\pm$ 62.22\\
$j-h$        &0.41$\pm$ 0.52&0.64$\pm$ 0.29\\
$h-k$        &0.13$\pm$ 0.37&0.61$\pm$ 0.37\\
$k+2.5logFpeak$&10.94$\pm$ 2.50&17.69$\pm$ 2.38\\
$k+2.5logFint$&11.16$\pm$ 2.56&17.85$\pm$ 2.46\\
$j+2.5logFpeak$&11.43$\pm$ 2.60&18.95$\pm$ 2.35\\
$j+2.5logFint$&11.70$\pm$ 2.65&19.10$\pm$ 2.42\\
\hline
\end{tabular}
\bigskip
\end{center}
\end{table*}

\begin{figure}[h!!!]

\includegraphics[bb=0 0 278 206,width=8cm,clip]{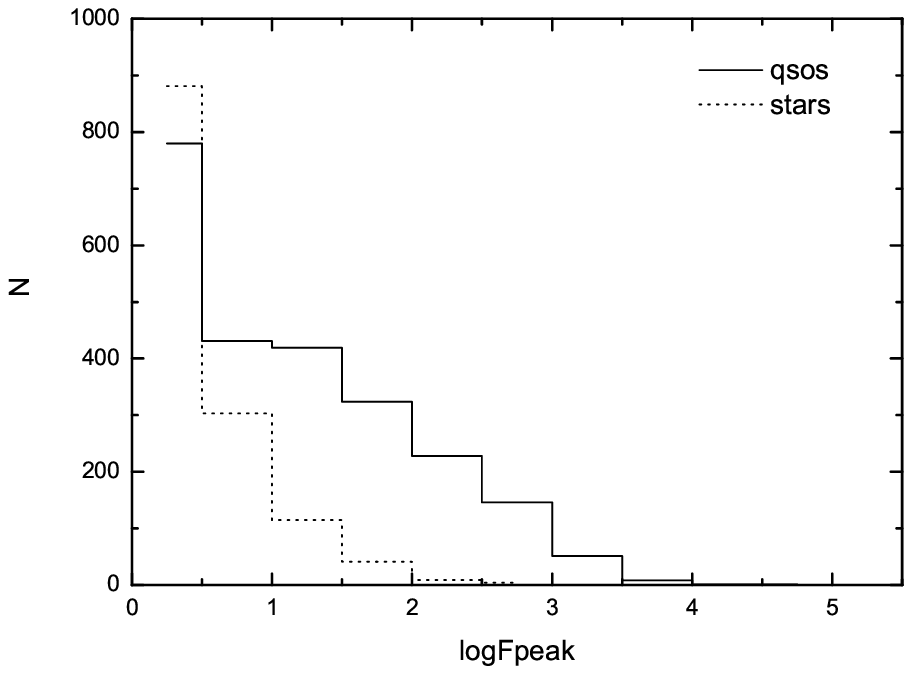}
\includegraphics[bb=0 0 278 206,width=8cm,clip]{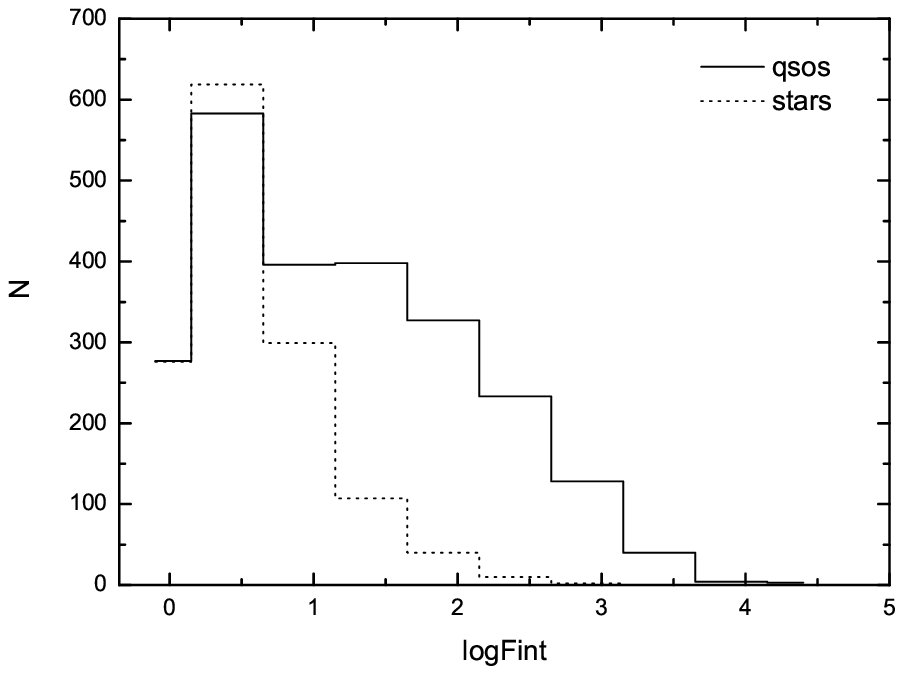}
\includegraphics[bb=0 0 278 206,width=8cm,clip]{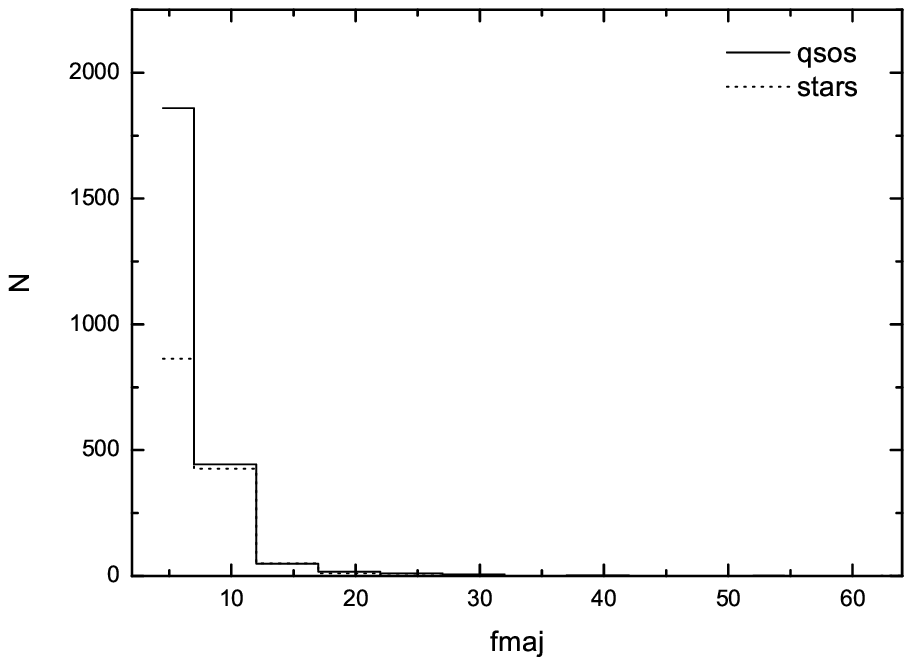}
\includegraphics[bb=0 0 278 206,width=8cm,clip]{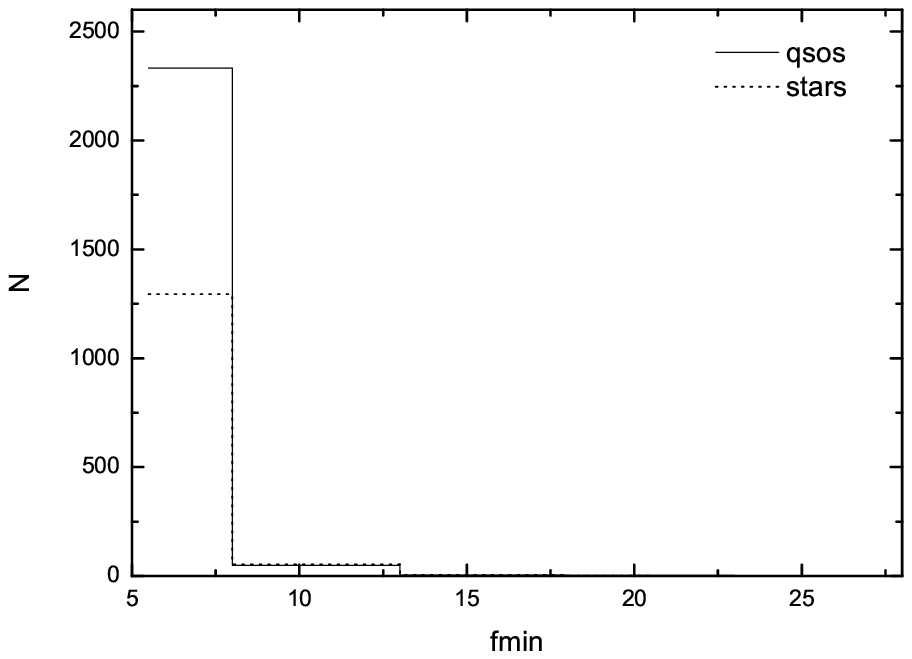}
\includegraphics[bb=0 0 278 206,width=8cm,clip]{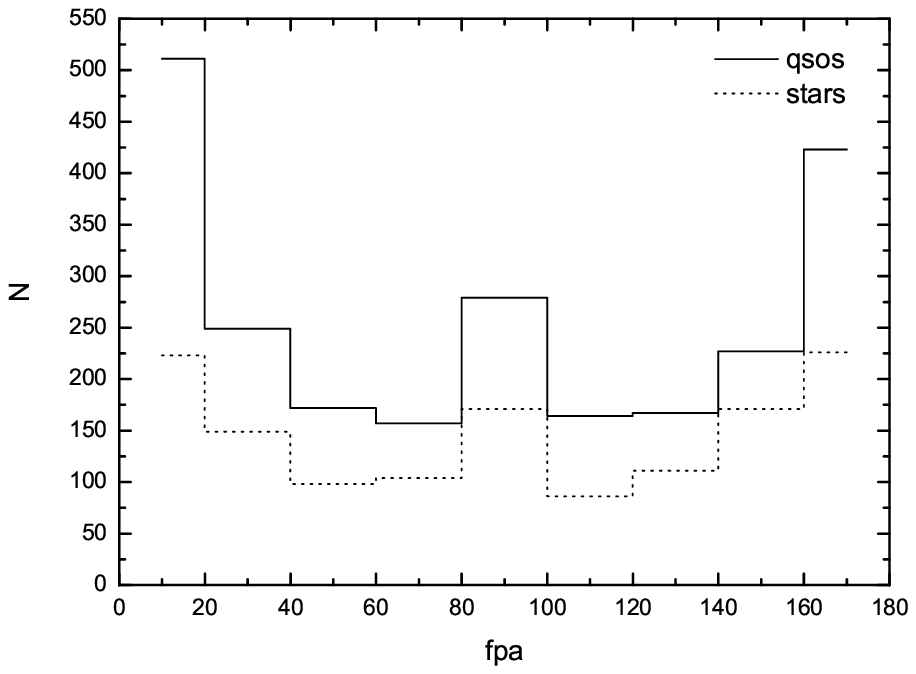}
\includegraphics[bb=0 0 278 206,width=8cm,clip]{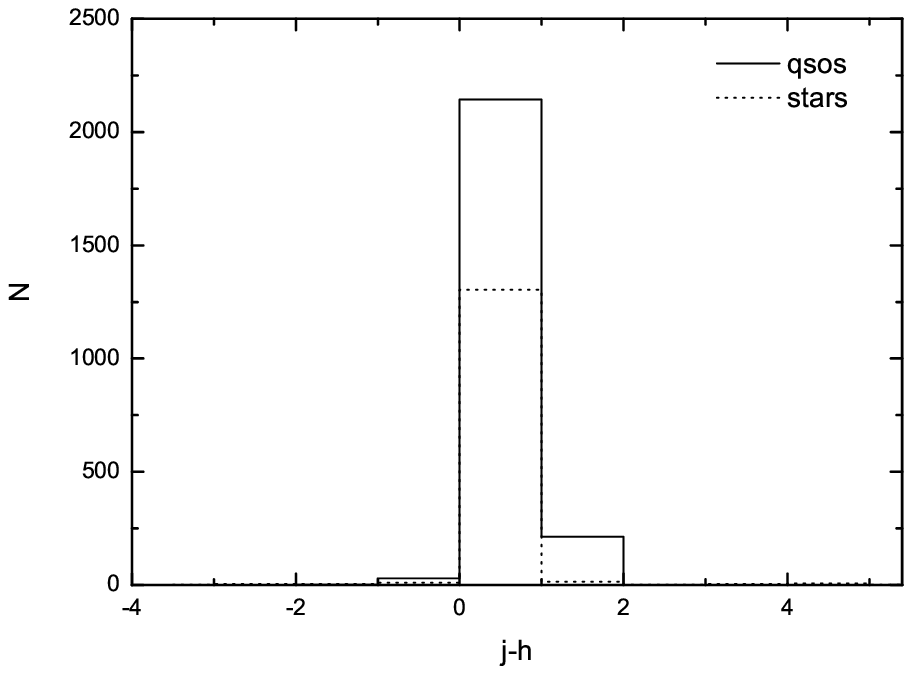}
\includegraphics[bb=0 0 278 206,width=8cm,clip]{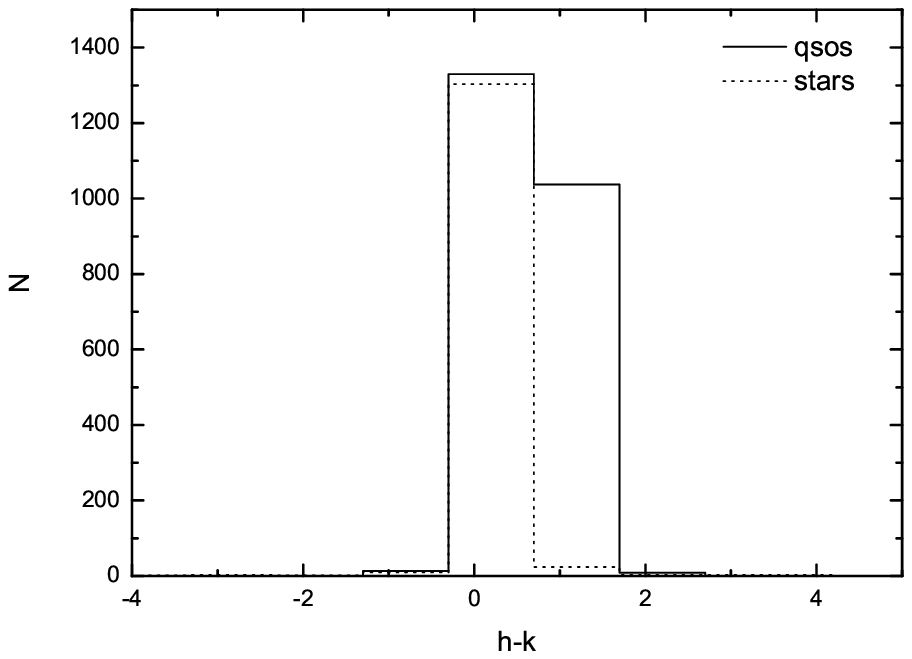}
\includegraphics[bb=0 0 278 206,width=8cm,clip]{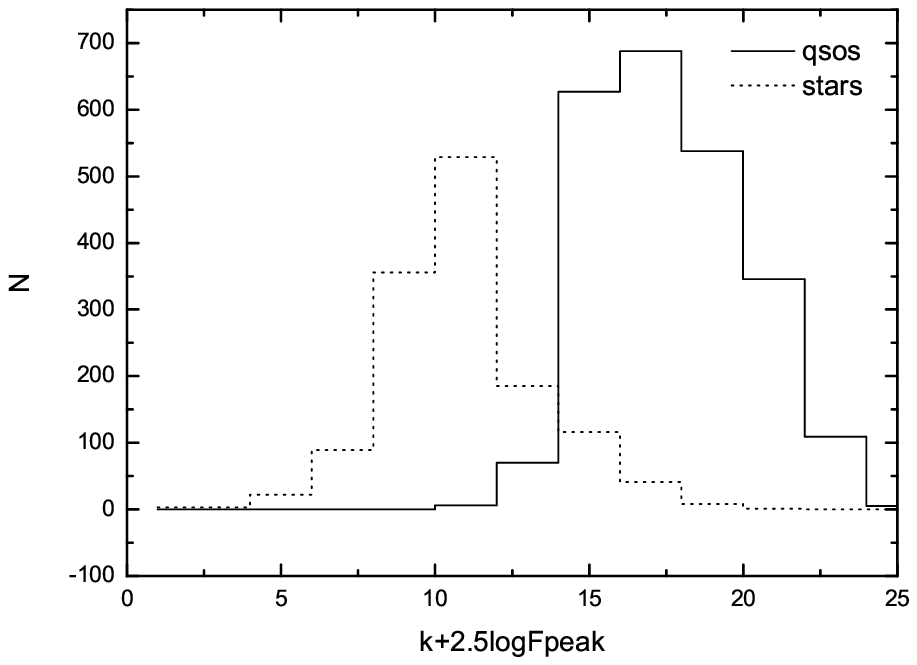}

\end{figure}

\begin{figure}[h!!!]
\includegraphics[bb=0 0 278 206,width=8cm,clip]{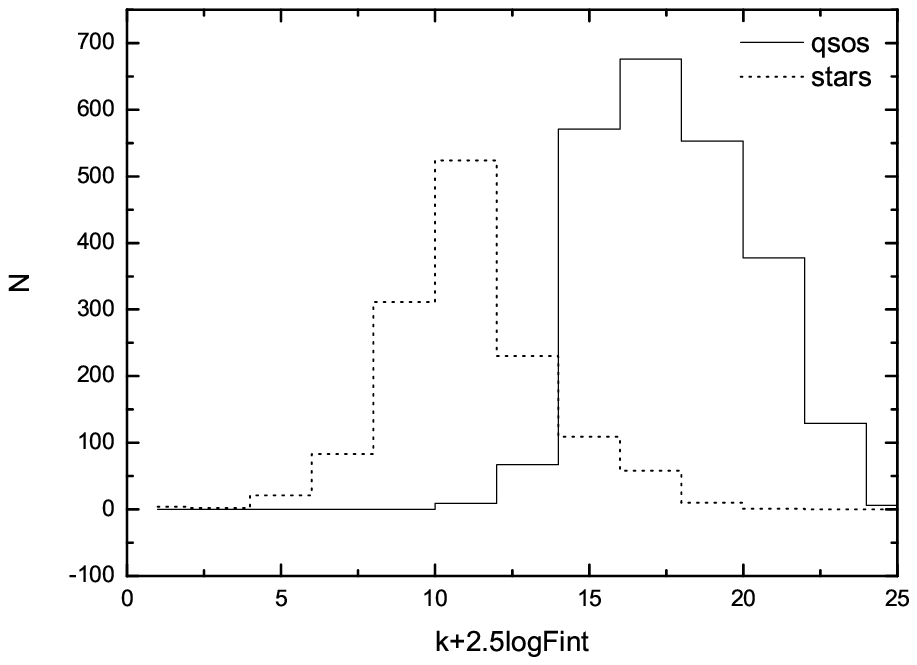}
\includegraphics[bb=0 0 278 206,width=8cm,clip]{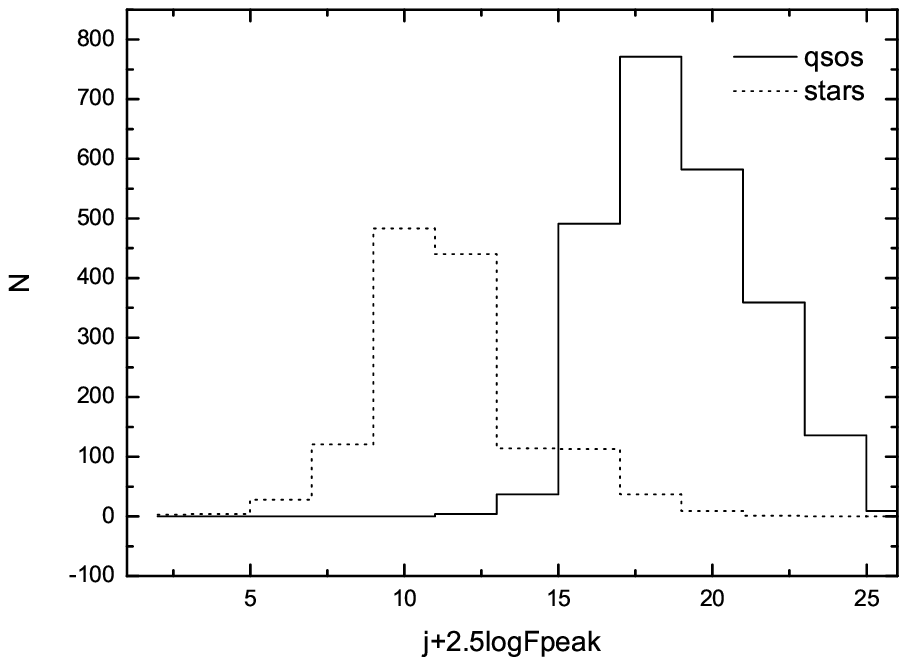}
\includegraphics[bb=0 0 278 206,width=8cm,clip]{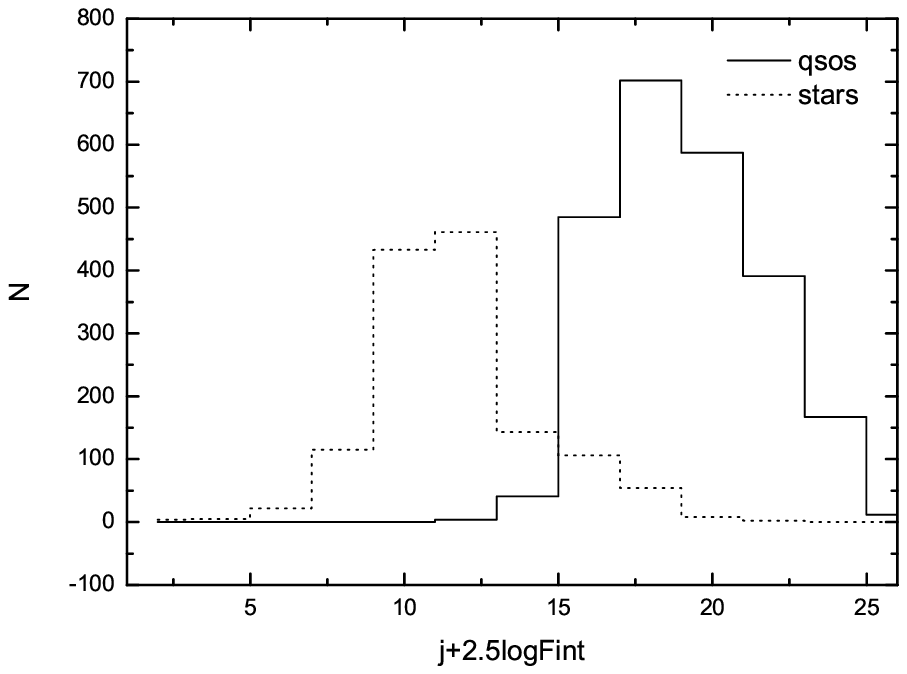}

\caption[]{Results of analysis of the sample for quasars (solid
line) and stars (dotted line).}
 \label{fig1}
\end{figure}

\section{Model Selection}
We used three methods to separate quasars from stars: Bayesian
Belief Networks (BBN), Multilayer Perceptron (MLP) networks and
Alternating Decision Trees (ADTree). BBN is used for the
classification of variable stars (L$\acute{o}$pez et al. 2006).
MLP and ADTree have been successfully used for the classification
of multiwavelength data, see Zhang et al. (2005), Zhang \& Zhao
(2004).

\subsection{Bayesian Belief Networks}
The Bayesian Belief Network (BBN) is a powerful knowledge
representation and reasoning tool under conditions of uncertainty.
BBN is defined by two components. The first is a direct acyclic
graph, where each node represents a random variable and each arc
represents a probabilistic dependence (Pearl 1988; Neapolitan
1990; Han \& Kamber 2001). If an arc is drawn from a node $Y$ to a
node $Z$, then $Y$ is a parent or immediate predecessor of $Z$,
and $Z$ is a descendent of $Y$. Each variable is conditionally
independent of its nondescendents in the graph, given its parents.
The variables may be discrete or continuous-valued. They may
correspond to actual attributes given in the data or to hidden
variables believed to form a relationship. The second consists of
one conditional probability table (CPT). The CPT for a variable
$Z$ specifies the conditional distribution $P(Z|Parent(Z))$, where
$parent(Z)$ are the parents of $Z$. The joint probability of any
tuple $(z_1,...,z_n)$ corresponding to the variables or attributes
$Z_1,...Z_n$ is computed by
$$
P(z_1,...,z_n)=\prod^{n}_{i=1}P(z_i|Parent(Z_i)),
$$
where the values for $P(z_i|Parents(Z_i))$ correspond to the
entries in the CPT for $Z_i$. A node within the network can be
selected as an output node, representing a class label attribute.
There may be more than one output node. Inference algorithms for
learning can be applied on the network. The classification
process, rather than returning a single class label, can return a
probability distribution for the class label attribute, that is
predicting the probability of each class.

\subsection{Multilayer Perceptron Networks}

The Multilayer Perceptron (MLP) network is one of the most widely
applied and investigated Artificial Neural Network model. MLP
networks have been applied successfully to solve some difficult
and diverse problems in training them in a supervised manner with
a highly popular algorithm, known as the error back-propagation
algorithm. The algorithm is based on the error-correction learning
rule. MLP network model consists of a network of processing
elements or nodes arranged in layers. Typically, it requires three
or more layers of processing nodes: an input layer which accepts
the input variables used in the classification procedure, one or
more hidden layers, and an output layer with one node for one
class. In fact, a network with just two hidden units using the
tanh function can fit the data quite well. The fit can be further
improved by adding yet more units to the hidden layer. However,
that having too large a hidden layer - or too many hidden layers -
can degrade the network's performance. In general, one shouldn't
use more hidden units than necessary to solve a given problem. One
way to ensure this is to start training with a very small network.
If gradient descent fails to find a satisfactory solution, grow
the network by adding a hidden unit, and repeat. MLP network is a
general-purpose, flexible, nonlinear model. Given enough hidden
units and enough data, it has been shown that MLPs can approximate
virtually any function to any desired accuracy. In other words,
any function can be expressed as a linear combination of tanh
functions: tanh is a universal basis function. Many functions form
a universal basis; the two classes of activation functions
commonly used in neural networks are the sigmoidal (S-shaped)
basis functions (to which tanh belongs), and the radial basis
functions. MLPs are valuable tools in problems when one has little
or no knowledge about the form of the relationship between input
vectors and their corresponding outputs. Examples of applications
of MLP networks in astronomy can be found in Vanzella et~al.
(2004). An introduction on Neural Networks is presented by Sarle
(1994a), and on multilayer Perceptron by Bailer-Jones et~al.
(2001) and Sarle (1994b). A comprehensive treatment of
feed-forward neural networks is provided by Bishop (1995).

\subsection{Alternating Decision Tree}

The alternating decision tree (ADTree) is a generalization of
decision trees, voted decision trees and voted decision stumps. A
general alternating tree defines a classification rule as follows.
An instance defines a set of paths in the alternating tree. As in
standard decision trees, when a path reaches a decision node it
continues with the child which corresponds to the outcome of the
decision associated with the node. However, when reaching a
prediction node, the path continues with all of the children of
the node. More precisely, the path splits into a set of paths,
each of which corresponds to one of the children of the prediction
node. We call the union of all the paths reached in this way for a
given instance the ``multi-path" associated with that instance.
The sign of the sum of all the prediction nodes which are included
in a multi-path is the classification which the tree associates
with the instance. The principle of the algorithm is explained in
Freund \& Mason (1999).

\section{Experiments and Results}

Our experiments were done with the WEKA machine learning package
(Witten \& Frank 2005). In the process of experimenting, the
default configurations of BBN, MLP and ADTree are used. The
computer used in this effort was a PC with a 3.4~GHZ Pentium 4 and
CPU 1~GB memory. The operating system was Microsoft Windows XP.
Here we use 10-fold cross-validation to evaluate the different
accuracy of different models for this database. By comparing the
accuracy of the classification and time taken to build models, we
try to compare the efficiency and effectiveness of the models.

\subsection{Cross-Validation}

Cross-validation is the statistical practice of partitioning a
sample of data into subsets such that the analysis is initially
performed on a single subset, while the other subset(s) are
retained for subsequent use in confirming and validating the
initial analysis. $K$-fold cross-validation is one important
cross-validation method. The data is divided into $k$ subsets of
(approximately) equal size. Each time, one of the $k$ subsets is
used as the test set and the other $k-1$ subsets are put together
to form a training set. Then the average error across all $k$
trials is computed. Cross-validation is often used for choosing
among various models, such as different network architectures. For
example, one might use cross-validation to choose the number of
hidden units, or one could use cross-validation to choose a subset
of the inputs (subset selection).

\subsection{Results}

Using the 10-fold cross-validation method, we found the
classification accuracy achieved with the different algorithms.
The results are shown in Tables 3-6. Here MLP employs a
three-layer topology, i.e. it includes one input layer, one hidden
layer and one output layer. Applying ADTree technique on the two
samples, the total number of nodes is 31 and the number of
predictor nodes is 21. For any algorithm, the accuracy of quasars
and stars is more than 88.0\%. Considering the sample 1 (S1),
correctly classified instances for BBN, MLP and ADTree are 3524,
3579 and 3553, respectively; as shown by Table 6, the
corresponding whole accuracy for BBN, MLP and ADTree amounts to
94.17\%, 95.64\% and 94.95\%, respectively; the running time to
build models is 0.34\,s, 25.14\,s and 1.25\,s, respectively.
Similarly, given the sample 2 (S2), correctly classified instances
for BBN, MLP and ADTree are 3531, 3585 and 3562, respectively;
Table 6 shows that the corresponding whole accuracy for BBN, MLP
and ADTree is 94.36\%, 95.80\% and 95.19\%, respectively; the
running time to build models is 0.28\,s, 19.23\,s and 0.86\,s,
respectively. From the results, we conclude that BBN, MLP and
ADTree are feasible to separate quasars from stars only
considering the accuracy. When only considering the running time,
BBN is the fastest, ADTree runs faster than MLP. If considering
both accuracy and time, ADTree is the best approach. Tables 3-6
also indicate that compared to the S1, the accuracy and the speed
to building the models for S2 all improve for different
algorithms. This fact clearly shows that the effectiveness and
efficiency of these models with feature selection are a little
better than those without feature selection. In addition, the
classification results indicate that it is applicable to preselect
quasar candidates from the 2MASS and FIRST survey catalogues. The
classifiers trained by these methods can be used to classify the
unclassified sources.

\begin{table*}[h!]
\begin{center}
\caption{The classification result for BBN with different samples}
\bigskip
\begin{tabular}{r|ll|ll}
\hline \hline
Sample&S1&&S2& \\
\hline
classified$\downarrow$known$\to$& stars &quasars& stars &quasars\\
\hline
        stars  &  1190  &  55&1190&48\\
       quasars &  163   &2334&163&2341\\
\hline
       Accuracy & 88.0\% & 97.7\%&88.0\%&98.0\% \\
\hline
\end{tabular}
\bigskip
\end{center}
\end{table*}

\newpage
\begin{table*}[ht]
\begin{center}
\caption{The classification result for MLP with different samples}
\bigskip
\begin{tabular}{r|ll|ll}
\hline \hline
Sample&S1&&S2& \\
\hline
classified$\downarrow$known$\to$& stars &quasars& stars &quasars\\
\hline
        stars    &   1220 &  30&1220&24\\
       quasars &   133   &2359&133&2365\\
\hline
       Accuracy & 90.0\% & 98.7\% &90.2\%&99.0\%\\
\hline
\end{tabular}
\bigskip
\end{center}
\end{table*}

\begin{table*}[ht]
\begin{center}
\caption{The classification result for ADTree with different
samples}
\bigskip
\begin{tabular}{r|ll|ll}
\hline \hline
Sample&S1&&S2& \\
\hline
classified$\downarrow$known$\to$& stars &quasars& stars &quasars\\
\hline
        stars    &  1194 &  30&1200&27\\
       quasars &   159   &2359&153&2362\\
\hline
       Accuracy &88.2\%  & 98.7\% &88.7\%&98.9\%\\
\hline
\end{tabular}
\bigskip
\end{center}
\end{table*}

\begin{table*}[h!]
\begin{center}
\caption{Accuracy and Time to built models for different methods
with different samples}
\bigskip
\begin{tabular}{r|ll|ll}
\hline \hline
Sample&S1&&S2& \\
\hline
Method& Accuracy & Time& Accuracy & Time\\
\hline
BBN & 94.17\%&0.34\,s&94.36\%&0.28\,s\\
MLP&95.64\%&25.14\,s&95.80\%&19.23\,s\\
ADTree& 94.95\%&1.25\,s&95.19\%&0.86\,s\\
\hline
\end{tabular}
\bigskip
\end{center}
\end{table*}

\section{Conclusions}

Survey data are one important source of information for
astronomers. By classification techniques, we can extract lots of
information from the raw data. Here we analyzed a sample and
compare the results with and without feature selection. When these
algorithms with feature selection are applied, the accuracy all
improves, and the speed to build models also accelerates,
comparing to the results without feature selection. Clearly
appropriate feature selection may improve the effectiveness and
efficiency of classifiers. For the given problem, BBN, MLP and
ADTree models on this sample achieve higher accuracy, more than
94.0\%. Only taking the accuracy into account, BBN, MLP and ADTree
performed comparably. But BBN classifier has fast speed when
applied to large databases, especially its speed is much faster
than those of MLP and ADTree. But in terms of both accuracy and
speed, ADTree shows its superiority. In conclusion, these
algorithms are robust and efficient methods for solving the
classification problems faced in astronomy. The classifiers
obtained by these methods may be used to preselect source
candidates in which astronomers are interested. These techniques
may be used on other types of astronomical data, such as spectral
data and image data. Moreover they are also applied on other
issues, for example star parameter measurement, redshift
estimation of galaxies and quasars, morphology classification of
galaxies. With the quantity, quality and complexity of
astronomical data improving and the number of features increasing,
selecting appropriate models and training the classifiers
efficiently, as well as feature selection methods, is a
challenging study for future research. Especially faced with large
and multiwavelength sky surveys, automated methods can not only
reduce astronomer's efforts, but also improve the efficiency of
astronomers and high-cost telescopes; moreover effective feature
selection methods reduce the dimensionality of space and improve
the efficiency and effectiveness of automated classification
algorithms, meanwhile they can make it possible for the
application of some methods only employed in low dimensional
spaces. The successful application of data mining in astronomical
databases is the catalyzer to find unusual, rare or unknown
objects and phenomenon. Especially clustering analysis and outlier
finding algorithms can facilitate class discovery in
astronomy.\\

\begin{acknowledgements}
We are very grateful to anonymous referee for his helpful comments
and suggestions. Meanwhile we show our thanks to Miss Gao Dan for
data processing. This research has made use of data products from
the Two Micron All Sky Survey, which is a joint project of the
University of Massachusetts and the Infrared Processing and
Analysis Center/California Institute of Technology, funded by the
National Aeronautics and Space Administration and the National
Science Foundation. This paper is funded by National Natural
Science Foundation of China under grant No.10473013 and
No.90412016.
\end{acknowledgements}

\label{lastpage}

\end{document}